\documentstyle[12pt]{article}
\newcommand{\be}{\begin{equation}}
\newcommand{\ee}{\end{equation}}
\newcommand{\bea}{\begin{eqnarray}}
\newcommand{\eea}{\end{eqnarray}}
\newcommand{\ba}{\begin{array}}
\newcommand{\ea}{\end{array}}

\begin{document}
\begin{titlepage}

\hspace{9.5cm}{YITP-P 99-54}

\vspace{.5cm}
\begin{center}
\LARGE

{\sc Susceptibility amplitude ratio near a Lifshitz point}

\vspace{.5cm}
\large

Marcelo M. Leite\renewcommand{\thefootnote}{\fnsymbol{footnote}}
\footnote[1]{e-mail: mmleite@insti.physics.sunysb.edu}
\renewcommand{\thefootnote}{arabic{footnote}}
\vspace{.5cm}

{\it C.N. Yang Institute for Theoretical Physics, SUNY at Stony Brook}
\\ {\it Stony Brook, NY 11794-3840, USA}

\vspace{.5cm}
October, 1999
\end{center}
\vspace{1cm}

\begin{abstract}
The susceptibility amplitude ratio in the neighborhood 
of a uniaxial Lifshitz point is calculated at one-loop level using 
field-theoretic and $\epsilon_{L}$-expansion methods. 
We use the Schwinger parametrization of the propagator in order to 
split the quadratic and quartic part of the momenta, as well as a new 
special symmetry point suitable for renormalization purposes. For a cubic 
lattice ($d = 3$), we find the result $\frac{C_{+}}{C_{-}} = 3.85$. 
\end{abstract}
\end{titlepage}
\newpage

\noindent

Universality is a key concept in the theory of critical phenomena, which 
states that all critical properties only depend on the number of components 
of the order parameter characterizing the phase transition and the 
space dimension of the system. Beside the critical exponents, the 
amplitude ratios above and below the critical temperature for different 
thermodynamic potentials are examples of universal quantities\cite{ref1}. One 
special type of critical behavior is associated with the Lifshitz 
point\cite{ref2}. In magnetic systems, the uniaxial Lifshitz point  
can be described by an axially nearest-neighbor Ising 
model(ANNI)\cite{ref3}, which consists of a spin-$\frac{1}{2}$ Ising model 
on a cubic lattice with nearest-neighbor ferromagnetic couplings and 
next-nearest-neighbor competing antiferromagnetic interactions along 
a single lattice axis. Due to the competition, the system presents a 
modulated phase (in addition to the ordinary para- and ferromagnetic 
ones). Theoretical and experimental studies in MnP\cite{ref4, ref5} 
showed that this system indeed presents this sort of uniaxial 
Lifshitz critical behavior.

Renormalization group and $\epsilon$-expansion techniques are 
particularly suitable to investigate amplitude ratios of critical 
systems\cite{ref6}. However, very little is known about these amplitude 
ratios for the Lifshitz critical behavior. The specific heat amplitude 
ratio for a uniaxial Lifshitz point was measured in MnP by Bindilatti, 
Becerra and Oliveira\cite{ref5}. 
Recently, some authors obtained this 
amplitude ratio theoretically at mean-field level\cite{ref7}. It turned out 
that the two results do not agree. This disagreement is not surprising, 
for the fluctuations must be taken into account in a proper treatment 
using the $\epsilon$-expansion. In order to find an outcome beyond 
mean-field for this amplitude ratio, one needs the coupling constant at 
two-loop level. As it is only known at one-loop for the Lifshitz point, 
we can then ask ourselves if it is possible to calculate 
some other amplitude ratio at one-loop order with this 
restricted knowledge of the coupling constant. If one considers the 
susceptibility amplitude ratio, such a program can be achieved. Besides, 
having a theoretical prediction for this amplitude ratio, where the 
renormalization group techniques can be exploited in its full power, should 
motivate experiments to test the degree of accuracy of this approach 
for systems of this type.

In this letter we calculate the susceptibility amplitude ratio at a Lifshitz 
point using $\lambda\phi^{4}$ field theory and $\epsilon$-expansion 
methods at first order in the loop expansion. In order to perform the 
one-loop integrals, we use 
the Schwinger parametrization for the free propagator, as well as a new 
special symmetry point. We will show that the result has the same dependence 
on $\epsilon_{L} = 4.5 - d$ for a uniaxial Lifshitz point as that exhibited 
by the usual Ising-like system, where the loop expansion parameter is 
$\epsilon = 4 - d$. We find the numerical value 
$\frac{C_{+}}{C_{-}} = 3.85$ for this amplitude ratio in a 
three-dimensional lattice. This is the first time that an amplitude ratio for 
the Lifshitz critical behavior is calculated to first order in $\epsilon_{L}$.

The most convenient way to formulate the $\lambda\phi^{4}$ field-theoretic 
approach to the Lifshitz point is the lagrangian description, which is 
equivalent to  the usual Landau-Ginzburg-Wilson hamiltonian formulation. For  
the uniaxial case, the bare lagrangian is:
\begin{equation}
L = \frac{1}{2}|\bigtriangledown_{1}^{2} \phi|^{2} + 
\frac{1}{2}|\bigtriangledown_{(d-1)} \phi|^{2} + 
\delta  \frac{1}{2}|\bigtriangledown_{1} \phi|^{2} 
+ \frac{1}{2} t_{0}\phi^{2} + \frac{1}{4!}\lambda\phi^{4} .
\end{equation}

We see that the competition along one axis produces the 
first term in the above expression. Furthermore, at the Lifshitz critical 
point 
$\delta = 0$. We are going to focus our attention in this case from now on.

The expression for the one-loop renormalized Helmholtz 
free-energy density at the fixed point associated with the uniaxial Lifshitz 
critical behavior of the system is:
\begin{eqnarray}
&& F(t,M) = \frac{1}{2} tM^{2} + \frac{1}{4!} g^{\ast} M^{4} + \frac{1}{4}
\left( t^{2} +  g^{\ast} t M^{2} + \frac{1}{4} g^{\ast^{2}} M^{4} \right) 
I_{sp} \nonumber \\
&& + \frac{1}{2}  \int d^{d-1} q dk
\{ \ln [ 1 + (1/2) g^{\ast} M^{2} / (k^{4} + q^{2}+ t) ] \nonumber \\
&& - \frac{1}{2} g^{\ast} M^{2} / (k^{4} + q^{2}) \}
\hspace{0.5cm}. 
\end{eqnarray}

\noindent
    
In the above equation $t, M \left( t_{0} = Z^{-1}_{\phi^{2}} t, \; 
\phi = Z^{-1/2}_{\phi} M \right)$ are the renormalized (bare) reduced
temperature and order parameter, respectively, $Z_{\phi^{2}}$,
$Z_{\phi}$ are renormalization functions, $g^{\ast}$ is the 
renormalized coupling constant  at the fixed point, $\vec{q}$ is 
a $(d-1)$-dimensional wave vector along the direction parallel to the plane 
where only ferromagnetic nearest-neighbor interactions take place, whereas 
$k$ is a wave vector parallel to the axis where the antiferromagnetic 
competition is localized. The integral $I_{sp}$ is defined by:
\begin{equation}
I_{sp} =  \int \frac{d^{d-1}q dk}{\left( (k + k^{'})^{4} + (q + p)^{2} 
\right) \left( k^{4} + q^{2}  \right)}.
\end{equation} 

\noindent

The symmetry point that simplifies the integral is chosen at external momenta 
$k^{'} = 0$, $p^{2} = 1$. This choice has the advantage of transforming the 
dimensionful coupling constant into the convenient dimensionless $u^{\ast}$. 
At this point is convenient to extract for each loop integration a geometric 
angular factor and absorb it in the coupling constant. In our case, it is 
$\frac{3\sqrt{2}}{8} S_{d-1}S_{1}$, where 
$S_{d} = [2^{d-1} \pi^{\frac{d}{2}} \Gamma(\frac{d}{2})]^{-1}$.  
In order to calculate this integral, we use the Schwinger parametrization:
\begin{eqnarray}
& & \int \frac{d^{d-1}{q}dk}{\left( k^{4} + (q + p)^{2} 
\right) \left( k^{4} + q^{2}  \right)} = \int^{\infty}_{0}\int^{\infty}_{0} 
d\alpha_{1}d\alpha_{2}
(2 \int^{\infty}_{0} dk exp(-(\alpha_{1} + \alpha_{2})k^{4})) \nonumber \\
& & \int d^{d-1}q exp(-(\alpha_{1} + \alpha_{2})q^{2} - 2\alpha_{2}q.p 
- \alpha_{2}p^{2}) .
\end{eqnarray}

The $q$ integral can be easily performed, 
\begin{eqnarray}
&& \int d^{d-1}q exp(-(\alpha_{1} + \alpha_{2})q^{2} - 2\alpha_{2}q.p 
- \alpha_{2}p^{2}) = \frac{1}{2} S_{d-1} \Gamma(\frac{d-1}{2})
(\alpha_{1} + \alpha_{2})^{- \frac{d-1}{2}} \nonumber \\ 
&& exp(- \frac{\alpha_{1} \alpha_{2}p^{2}}{\alpha_{1} + \alpha_{2}}),
\end{eqnarray}
and the $k$ integration is \cite{ref8} :

\begin{equation}
2 \int^{\infty}_{0} dk exp(-(\alpha_{1} + \alpha_{2})k^{4}) = 
\frac{1}{2}(\alpha_{1} + \alpha_{2})^{- \frac{1}{4}} \Gamma(\frac{1}{4}).
\end{equation}

Replacing equations (5), (6) into equation (4) together with the value 
$p^{2}=1$, one finds
\begin{eqnarray}
&& \left(\int \frac{d^{d-1}{q}dk}{\left( k^{4} + (q + p)^{2} 
\right) \left( k^{4} + q^{2}  \right)}\right)_{p^{2} = 1} =  
\frac{1}{4}S_{d-1} \Gamma(\frac{d-1}{2}) \Gamma(\frac{1}{4}) \nonumber \\ 
&& \int^{\infty}_{0} \int^{\infty}_{0} d\alpha_{1}d\alpha_{2} 
exp(- \frac{\alpha_{1} \alpha_{2}}{\alpha_{1} + \alpha_{2}})
(\alpha_{1} + \alpha_{2})^{-(\frac{d-1}{2} + \frac{1}{4})}.
\end{eqnarray}

We can perform one of the integrals in the Schwinger parameters using 
a change of variables. Then, after a rescale, the result can 
be expressed in the following form\cite{ref9} 
\begin{eqnarray}
&& \int^{\infty}_{0} \int^{\infty}_{0} d\alpha_{1}d\alpha_{2} 
exp(- \frac{\alpha_{1} \alpha_{2}}{\alpha_{1} + \alpha_{2}})
(\alpha_{1} + \alpha_{2})^{-(\frac{d-1}{2} + \frac{1}{4})} = 
\Gamma(2 - (\frac{d-1}{2} + \frac{1}{4})) \nonumber \\
&& \int^{1}_{0} dv(v(1-v))^{(\frac{d-1}{2} + \frac{1}{4}) - 2} .
\end{eqnarray}
 
Now we make the continuation $d = 4.5 - \epsilon_{L}$. We can make use of 
the identity 
$\Gamma(1.75 - \frac{\epsilon_{L}}{2})\Gamma(0.25) = \frac{3\sqrt{2}\pi}{4}
\Gamma(2 - \frac{\epsilon_{L}}{2})$, to get the following expression for 
$I_{sp}$
\begin{equation}
I_{sp} = \frac{1}{\epsilon_{L}}(1 + \frac{\epsilon_{L}}{2}) .
\end{equation}

We are now in position to calculate the susceptibility amplitude ratio. Using 
equation (2) we find the following renormalized equation of state :  
\begin{eqnarray}
H_{R} & = & \frac{\partial F}{\partial M} = tM + \frac{1}{6} u^{\ast} 
M^{3} + \frac{1}{2} u^{\ast} M \left( t + \frac{1}{2} u^{\ast} M^{2} \right)
\nonumber \\
& \times & \left[ I_{sp} -  \int \frac{d^{d-1}q dk}{\left(
k^{4} + q^{2}  \right) \left( k^{4} + q^{2}  + t + \frac{1}{2}
u^{\ast} M^{2} \right)} \right] \hspace{0.5cm} .
\end{eqnarray} 

The one-loop integral is then readily calculated:
\begin{equation}
\int \frac{d^{d-1}{q} dk}{\left( k^{4} + q^{2}
\right) \left( k^{4} + q^{2}  + t + \frac{1}{2}
u^{\ast} M^{2} \right)}  = \frac{1}{2} \Gamma(2 - \frac{\epsilon_{L}}{2})
\Gamma(\frac{\epsilon_{L}}{2})  
(t + \frac{1}{2}u^{\ast} M^{2})^{-\frac{\epsilon_{L}}{2}}.
\end{equation}

The renormalized two-point vertex part
\begin{equation} 
\Gamma^{(2,0)}_{R} = \frac{\partial}{\partial M} H_{R},
\end{equation} 
is related to the susceptibility as 
\begin{equation}
\chi^{-1} =\Gamma_{R}^{(2,0)}.
\end{equation} 

We can now apply the following procedure to calculate this 
amplitude ratio\cite{ref10}. For $T > T_{L}$ we can put $M = 0$ 
into equation (12) above and use the Lifshitz value at the fixed point 
$u^{\ast} = \frac{2\epsilon_{L}}{3}$, to get
\begin{equation}
\chi( T > T_{L}) = t^{- \gamma_{L}}(1 - \frac{\epsilon_{L}}{6}).
\end{equation} 
 
For $T < T_{L}$, we use $u^{\ast}M^{2} = -6t$ and proceeding along the same 
lines gives the result
\begin{equation}
\chi(T < T_{L}) = (-t)^{-\gamma_{L}} \frac{1}{2}
(1 - \frac{\epsilon_{L}}{6}(4 + ln2)) ,
\end{equation}
with amplitude ratio
\begin{equation}
\frac{C_{+}}{C_{-}} = 2^{\gamma_{L} -1} \frac{\gamma_{L}}{\beta_{L}} , 
\end{equation}
where $\gamma_{L} = 1 + \frac{\epsilon_{L}}{6}$ and 
$\beta_{L} = \frac{1}{2} - \frac{\epsilon_{L}}{6}$ are the susceptibility 
and magnetization critical exponents, respectively, associated to 
the Lifshitz point. First, we note that expression (16) has the same 
dependence on $\epsilon_{L}$ as the usual Ising-like critical behavior, 
the only difference being the value of $\epsilon_{L} = 1.5$ for a 
cubic lattice($d = 3$). The numerical value for the amplitude ratio is 
then $\frac{C_{+}}{C_{-}} = 3.85$. Compared with the value 
$\left(\frac{C_{+}}{C_{-}}\right)_{mean-field} = 2$, the correction due to the 
fluctuations is remarkable. Second, the 
method developed here might be efficient to calculate the fixed point at 
two-loop level, and then to find the specific heat amplitude ratio at order 
$\epsilon_{L}$ in order to compare with known experimental data\cite{ref5}. 
Alternatively, the result obtained for the susceptibility amplitude 
ratio should motivate the realization of experiments to check whether the 
renormalization group techniques are suitable to understand this 
sort of system. Indeed, the comparison of the critical exponents $\beta_{L}$ 
and $\gamma_{L}$ to first order in $\epsilon_{L}$ with Monte Carlo 
simulations showed that they are different\cite{ref3}. It was argued that 
the Monte Carlo result was more appropriate, because the expansion 
parameter $\epsilon_{L}$ is not small and, therefore, the perturbative 
expansion might not be reliable. On the other hand, carrying out the 
calculation of the critical exponents to second order in $\epsilon_{L}$, 
might actually bring their values closer to those obtained via Monte Carlo. 
The definite answer to either possibility has to wait until one can figure out 
the fixed point at two-loop order. As Monte Carlo methods are not available 
yet to calculate amplitude ratios, the most direct way to 
probe the numerical value at order $\epsilon_{L}$ of the susceptibility 
amplitude ratio shown here is to compare with experiments 
to be done in systems with uniaxial critical Lifshitz behavior, such as MnP. 
This comparison should give a clue about the reliability of the 
$\epsilon_{L}$ expansion methods in this case. 
Finally, although some authors have recently proposed a different 
field-theoretic approach to the Lifshitz point\cite{ref11}, their method 
does not seem to be suitable for the uniaxial case, for their choice 
of the symmetry point makes the integral $I_{sp}$ more difficult to 
be performed. We hope to discuss the issues of crossover and two-loop 
calculations elsewhere.    

\large {\bf Acknowledgements}
\normalsize

The author would like to thank Denis Dalmazi for useful discussions and for 
pointing out reference \cite{ref9}. Support from FAPESP, grant number 
98/06612-6, is gratefully acknowleged.

\end{document}